\documentclass[aps,prl,reprint,superscriptaddress
]{revtex4-2}
\usepackage{graphicx}
\usepackage{dcolumn}
\usepackage{rotating}
\usepackage{amssymb}
\usepackage{mathptmx}
\usepackage{amsfonts}
\usepackage{amsmath}
\usepackage{xcolor}
\usepackage{bm}
\usepackage{gensymb}
\usepackage[colorlinks=true,linkcolor=blue,urlcolor=blue,citecolor=blue]{hyperref}
\usepackage{dcolumn}% Align table columns on decimal point
\usepackage{bm}% bold math
\usepackage{color}
\newcommand{\vso}{$\beta$--VOSO$_4$}

\begin{document}

\preprint{}

\title{One-dimensional magnetism in synthetic Pauflerite, \vso}

\author{Diana Lucia Quintero-Castro}
\email[]{diana.l.quintero@uis.no}
\affiliation{Department of Mathematics and Physics, University of Stavanger, 4036 Stavanger, Norway}%
\affiliation{Helmholtz Zentrum Berlin f\"ur Materialien und Energie, D-14109 Berlin, Germany}

\author{G\o{}ran J. Nilsen}%
\affiliation{ISIS Neutron and Muon Source, STFC Rutherford Appleton Laboratory, Didcot OX11 0QX, United Kingdom
}%
\affiliation{Department of Mathematics and Physics, University of Stavanger, 4036 Stavanger, Norway}%

\author{Katrin Meier-Kirchner}
\affiliation{Helmholtz Zentrum Berlin f\"ur Materialien und Energie, D-14109 Berlin, Germany}

\author{Angelica Benitez-Castro}
\affiliation{Laboratory for Magnetism and Advanced Materials, Universidad Nacional de Colombia, Manizales 17003, Colombia}

\author{Gerrit Guenther}
\affiliation{Helmholtz Zentrum Berlin f\"ur Materialien und Energie, D-14109 Berlin, Germany}

\author{Toshiro Sakakibara}
\affiliation{The Institute for Solid State Physics, The University of Tokyo, Kashiwa, Chiba 277-8581, Japan}

\author{Masashi Tokunaga}
\affiliation{The Institute for Solid State Physics, The University of Tokyo, Kashiwa, Chiba 277-8581, Japan}

\author{Chidozie Agu}
\affiliation{Department of Mathematics and Physics, University of Stavanger, 4036 Stavanger, Norway}

\author{Ipsita Mandal}
\affiliation{Department of Mathematics and Physics, University of Stavanger, 4036 Stavanger, Norway}
\affiliation{Institute of Nuclear Physics, Polish Academy of Sciences, 31-342 Krak\'{o}w, Poland}

\author{Alexander A. Tsirlin}
\affiliation{Felix Bloch Institute for Solid-State Physics, Leipzig University, 04103 Leipzig, Germany}

\date{\today}

\begin{abstract}

We have synthesized single-crystal samples of \vso \, and fully characterized their magnetic properties. Our magnetic susceptibility, high field magnetization and powder inelastic neutron scattering results are in excellent agreement with theoretical expressions for a one-dimensional spin-1/2 Heisenberg chain with an exchange parameter of $3.83(2)$\,meV. \emph{Ab initio} calculations identify the superexchange pathway, revealing that the spin-chain does not run along the expected crystallographic chain $a$ direction but instead between V$^{4+}$O$_{6}$ octahedra that are linked via SO$_{4}$ tetrahedra along the $b$ axis. We do not detect any phase transition to a long-range magnetic order within our experimental conditions, indicating \vso \, is very close to an ideal one-dimensional magnetic system. 

\end{abstract}

\maketitle

\section{Introduction}

Low-dimensional Heisenberg systems are a fertile testing ground which offer well-recognised model materials \cite{MotoyamaN1996Msoi, TennantD.Alan2005Qcau}, exact analytical solutions \cite{BetheH.1931ZTdM, PhysRevB.55.8894}, multiple experimental predictions \cite{Caux2006, PhysRevLett.128.187202}, newly proposed quantum simulations \cite{PhysRevResearch.4.013193} and technological applications, such as quantum spin transistors \cite{MarchukovO.V.2016Qstw}. Many low-dimensional Heisenberg magnets are $d^1$ V$^{4+}$ spin$-1/2$ (s$-1/2$) oxides. In these compounds, the crystal field environment, with a distinctive short V-O bond, lifts the degeneracy of the t$_{2g}$ orbitals, hence quenching the orbital magnetic moment. The lowest-energy half-filled $d_{xy}$ orbital lies in the plane perpendicular to the short bond, where it can overlap with oxygen orbitals, creating superexchange paths that give rise to a dominant planar antiferromagnetic (AFM) interaction, and hence to low-dimensional spin systems \cite{Savina, tsirlin2011, tsirlin2011b, arjun2019, mazurenko2006, korotin1999}. 

\begin{figure}[htb!]
\includegraphics[width=0.45\textwidth]{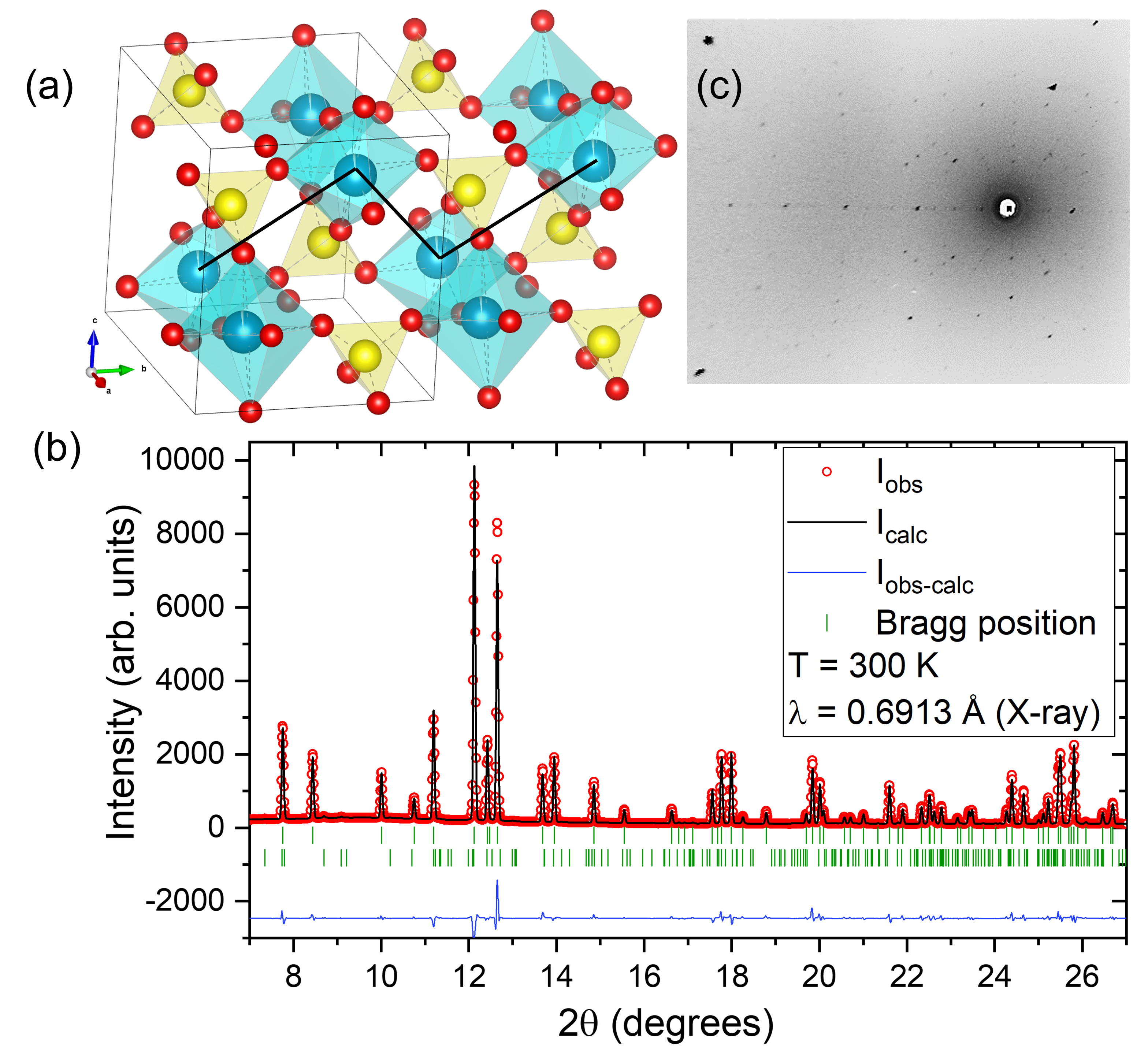}
\caption{(a) Graphic representation of the crystal structure of \vso \, created using \texttt{VESTA} \cite{vesta}. SO$_{4}$ tetrahedra are represented in yellow, while V$^{4+}$O$_{6}$ octahedra are blue showing a partial occupation of the V$^{4+}$ site following Ref \cite{Fuller}. Oxygen ions are in red. The black line represents the main magnetic exchange interaction as described in the sessions below. (b) X-ray diffraction pattern of \vso \, at $300$\,K along with its Rietveld refinement. Experimental data in red, calculated pattern in black and difference in blue. There are two sets of Bragg positions -above- the positions related to the \vso \, structure and the ones below are due to the VH$_6$SO$_8$ impurity. (c) X-ray Laue image with the beam parallel to the c-axis for the sample used for magnetometry. 
}
\label{fig:AtomicStruc}
\end{figure}

\vso \, was first synthesized in 1928 by Sieverts and M\"uller \cite{Sieverts} by the reduction of  V$^{5+}$ in sulfuric acid. Later, in 2007, it was discovered among volcanic products of the Tolbachik volcano in Kamchatka, Russia \cite{KrivovichevSergeyV.2007PbVs}.
Vanadyl based compounds like \vso \, and its hydro-oxosulfate modifications attract the interest of chemists and biologists due to their potential applications in catalysis and the management of diabetes \cite{Goc+2006+314+332, catalyst}. 
\vso \, crystallizes in the $Pnma$ orthorhombic group and its structure represents a three-dimensional framework built up of distorted  V$^{4+}$O$_{6}$ octahedra and (SO$_{4}$)$^{2-}$ tetrahedra. Recent crystallographic studies have revealed 2\% of correlated defects in the here investigated synthetic crystals \cite{Fuller}. These are linked to an inversion of the short-long V-O distance pairs and form thin layers with a negative inter-layer correlation, randomly destroying the alternation of the V-O bonding pattern along the $b$ axis. Figure \ref{fig:AtomicStruc}(a) shows a representation of the crystal structure of \vso \,following Ref.\cite{Fuller}  showing a $98\%$ occupation of the vanadium site at $(0.334, 0.25, 0.2675)$ .

Previous studies of the magnetism in \vso \, through magnetic susceptibility measurements report long-range antiferromagnetic order \cite{LONGO1970394} and  ferrimagnetic behaviour \cite{ferrimag}. Here, we revise these previously published interpretations in light of data from newly grown mm-sized single crystals. We characterize the magnetic properties, through magnetic susceptibility, high field magnetization measurements and powder inelastic neutron scattering (INS). The INS data shows clear gapless spinons with a continuum spanning up to 12\,meV. These data are compared to calculations of the  spinon continuum of the one-dimensional (1D) Heisenberg chain. Reasonable agreement is found for $J_1=3.83(2)$~meV, with the dominant exchange running along the $b$-direction within the exchange path predicted by our $ab$-initio calculations. We find the effect of interchain couplings to be negligible and confirm the system to be a 1D s-$1/2$ Heisenberg chain magnet. 

\section{Experimental and Calculation Details}

Single crystal samples of \vso \, were synthesized through the reaction: V$_2$O$_5$+2H$_2$SO$_4$ $\rightarrow$ 2\vso + 2H$_2$O + 0.5O$_2$, following Ref.\cite{Sieverts}. A solution of 3\,g of V$_2$O$_5$ and 100\,ml H$_2$SO$_4$ was heated up just below the boiling point of H$_2$SO$_4$ at 290$^\circ$C in a conical flask for a maximum period of $3$ months. A fractionating column was used to condense all vapors. Slow heating and cooling rates of 135$^\circ$C/h and 30$^\circ$C/h were used, respectively.  All products were washed multiple times in an ice-water bath and dried at 110$^\circ$C for 12 h. Thermogravimetric analysis was performed to ensure no water was left in the samples. These results (not shown here) were in agreement with those of Paufler $et$ $al.$, \cite{Paufler}. Our longest reaction of 3 months allowed the growth of needle-like single crystals as long as 6\,mm of a green colour. X-ray Laue diffraction [Fig. \ref{fig:AtomicStruc}(c)] was used to confirm the crystallinity of the samples. Powder samples were produced by grinding the single crystals.
Indications of a small amount of the impurity phase VH$_6$SO$_8$ (space group $P2_1/c$) were found in X-ray powder diffraction data. Figure \ref{fig:AtomicStruc}(b) shows the X-ray powder diffraction data along with its refinement. This data was acquired using the BM01 diffractometer at SNBL-ESRF using a wavelength of $0.6913$\,$\rm{\AA}$ . The impurity is contained in the powder matrix where the crystals grow (not in the single crystals). The content varies in the powder sample but it amounts to less than $1\%$ of the total mass. 

A single crystal sample of mass $7.021$\,mg was used to measure magnetic susceptibility using a Quantum Design superconducting quantum interference device magnetometer. These measurements were carried out in both zero-field and field-cooled modes  with a magnetic field of $0.5$\,T applied along the three main crystallographic directions. The temperature range for the measurements was $2–300$\,K. This magnetometer was also used to measure magnetization in a single crystal of mass $27.97$\,mg with magnetic fields up to 5\,T applied along the $a$ axis at 4.2\,K. This measurement was performed in order to calibrate the magnetic moment measured in the pulsed high magnetic fields using an induction method at the International MegaGauss Science Laboratory. This measurement was also performed at 4.2\,K on the same sample and fields up to 66.3\,T were achieved.
Additionally, a home-made high sensitivity magnetometer was used to measure the magnetic susceptibility in the temperature range $0.085$ and $2.5$\,K with a magnetic field of $0.5$\,T applied along the $c$ axis \cite{shimizu}.

Inelastic neutron scattering (INS) experiments were performed on a polycrystalline sample of mass 2.614\,g in an annular shape Al-foil holder, at 3\,K using the direct geometry neutron ToF instrument NEAT at HZB \cite{neat}. Measurements were performed with neutrons of incident wavelengths of 2, 3 and 5\,$\rm{\AA}$ (incident energies $20.45$, $9.09$ and $3.27$\,meV respectively).

Density-functional (DFT) band-structure calculations were performed in the \texttt{FPLO} code \cite{fplo} using the generalized gradient approximation (GGA) for the exchange correlation potential \cite{pbe96}. A dense \textbf{k}-mesh with up to 343 points in the symmetry-irreducible part of the Brillouin zone was used.

\section{Results and Calculations}

\subsection{Static susceptibility and magnetization}

\begin{figure}[htb!]
\includegraphics[width=0.52\textwidth]{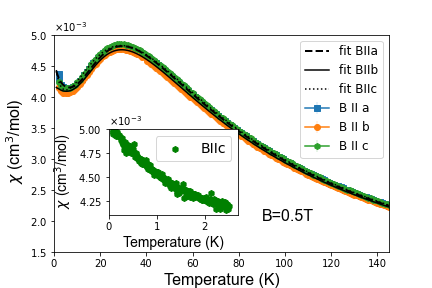} 
\caption{Static susceptibility data for a magnetic field of $0.5$\,T applied along each main crystallographic direction, displayed along with the fitted
expression for a $s-1/2$ uniform Heisenberg chain given by eq.~$50$ in Ref. \cite{Johnston}. Details of the low-temperature range (85\,mK to 2.5\,K) with field applied along the $c$ axis are shown in the inset.}
\label{fig:susceptibility}
\end{figure}

The magnetic static susceptibility of \vso \, with magnetic field of $0.5$\,T applied along the main three crystallographic axes is shown in Fig.~\ref{fig:susceptibility}. The three curves are slightly different, suggesting a  weakly anisotropic \emph{g} tensor. The Curie-Weiss temperatures and Curie constants obtained from a fit to the Curie-Weiss law above 150\,K are listed in Table \ref{table:suscept}. The former temperatures are between $-34.5$\,K and $-39.4$\,K, suggesting dominant AFM interactions.  
Zero field cooled and field cooled measurements were done, and showed no sign of any hysteresis processes in the measured ranges (not shown here). All three susceptibility plots show a broad maximum centred at $\approx29$\,K. Such a susceptibility curve is characteristic of low-dimensional magnets. 
The inset in Fig.~\ref{fig:susceptibility} shows the magnetic susceptibility for magnetic fields applied along the $c$ axis and temperatures down to $0.085$\,K. No sign of a transition to a long-range magnetic ordered state was detected within this experimental range. 

Assuming the hypothesis that \vso \, is an s$-1/2$ antiferromagnetic Heisenberg chain system, we use the expression calculated by Johnston, \emph{et al.}, \cite{Johnston} [Eqs.~30 and 31], for the characteristic temperature of the maximum of the magnetic susceptibility T$_{\chi_{max}} = 0.640851|J_{1D}|/k_B$. From this expression, we can estimate an intra-chain interaction ($J_{1D}$) of $3.9$\,meV for fields along the $a$ and $c$-directions and $3.78$\,meV for fields along the $b$ axis. Furthermore, by setting the $g$ factors to those reported in Ref. \cite{LAGUTA2019228}, we obtain $\chi_{max}$T$_{\chi_{max}}=0.035477 g^2\frac{cm^3K}{mol}$ for fields along the $a$ axis, $\chi_{max}$T$_{\chi_{max}}=0.037857 g^2\frac{cm^3K}{mol}$ for the $b$ axis and $\chi_{max}$T$_{\chi_{max}}=0.035744 g^2\frac{cm^3K}{mol}$ for the $c$ axis. These values are very close to the theoretical value for a $s-1/2$ AF uniform Heisenberg chain where $\chi_{max}$T$_{\chi_{max}}=0.0353229(3) g^2\frac{cm^3K}{mol}$ \cite{Johnston}. 

Giving these similarities, we proceed to fit the data to the mean-field approximation for the magnetic susceptibility of an s$-1/2$ uniform ($\alpha=1$) Heisenberg chain with an intra-chain interaction ($J_{1D}$), following eq.~$50$ in Ref. \cite{Johnston}. The total fitting function also includes a Curie Weiss law term to account for the contribution of paramagnetic impurities $[C_{imp}/(T-\theta_{CW-imp})]$ which dominate the data below $7$\,K and a background susceptibility ($\chi_0$) to account for core diamagnetism and van Vleck contributions. All resultant parameters are listed in Table \ref{table:suscept}. The impurity Curie constant amounts to 1-1.45\% of the \vso \, Curie constant values. These paramagnetic impurities could be well linked to the 2\% defect planes reported in Ref. \cite{Fuller}. 
The \emph{g} values were also extracted using the fit to eq.~$50$ in Ref. \cite{Johnston}. These values are very close to 2 but are slightly bigger than those typically reported for V$^{4+}$ \cite{EPR}. In comparison the g-values extracted from the $\chi_{max}$T$_{\chi_{max}}$ expressions  are ${1.991, 1.947, 1.998}$ for fields parallel to each main crystallographic axis.

\begin{table*}[!htb]
\caption{Parameters obtained by fitting the susceptibility data both to the Curie-Weiss law (above 150\,K and resulting C and $\theta_{CW}$) and to eq.~50 in Ref. \cite{Johnston} in the whole temperature range.}
\label{table:suscept}
\renewcommand{\arraystretch}{1}
\begin{ruledtabular}
\begin{tabular}{c  c c c  c  c c c c } 
 & C (cm$^3$/mol K)  & $\theta_{CW}$ (K) & J$_{1D}$ (meV) & g &  C$_{imp}$ (cm$^3$/mol K) &  $\theta_{CW-imp}$ (K) & $\chi_0$ (cm$^3$/mol) 
\tabularnewline
\hline
B $\parallel a$ & 0.4034(3) & -34.5(2) &  4.107(5) & 2.006(7) & 0.004(1) & -3.8(2) & $3.6(1)\times10^{-5}$ 
   \tabularnewline
B $\parallel b$ & 0.4103(2) & -39.4(2) &  4.108(6) & 1.979(1) & 0.006(4) & -7.9(5) & $6.4(1)\times10^{-5}$ 
   \tabularnewline
B $\parallel c$ & 0.4056(4) & -35.0(3) &  4.11(5) & 2.010(7)  & 0.005(2) & -6.8(4) & $3.1(1)\times10^{-5}$  
  \tabularnewline
\end{tabular}
\end{ruledtabular}
\end{table*}

\begin{figure}[htb!]
\includegraphics[width=0.45\textwidth]{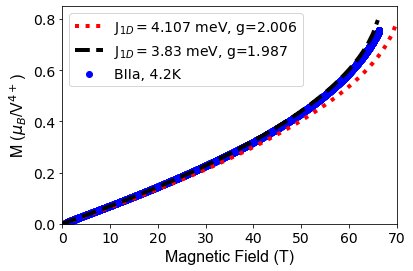}
\caption{Magnetization per V$^{4+}$ ions as a
function of applied magnetic field,
performed using a pulsed magnet at
4.2\,K and field applied parallel to the $a$ axis (blue). The red dashed line corresponds to eq.~\ref{eq:magn} using the magnetic exchange interactions obtained from the susceptibility data without any fitting. The black dashed line corresponds to the calculated using the same equation but with the magnetic exchange interaction derived from the fit to the inelastic neutron scattering data and the published g factor for V$^{4+}$ \cite{EPR}.}
\label{fig:magn}
\end{figure}

The high field magnetization of \vso \, at 4.2\,K for fields applied along the $a$ axis is shown in Fig.~\ref{fig:magn}. The magnetization increases with $H$ following the function:

\begin{align} \label{eq:magn}
M=M_{sat}(1-\sqrt{1-(H/H_{sat})})
\end{align}

\noindent where $M_{sat}$ is the saturation magnetization per V$^{4+}$ ion, H is magnetic field and $H_{sat}$ is the saturation field, $68.46$\,T, following the description given in Ref.~\cite{Bonner}. The red dotted line in  Fig.~\ref{fig:magn} corresponds to this equation with $J_{1D}=4.107$\,meV and $g=2.006$ obtained by the susceptibility fit for the $a$ axis. No fitting procedure was done in the case of the magnetization modeling. The magnetic exchange interaction extracted from the susceptibility follows well the magnetization below $20$\,T after which it deviates given a much larger saturation field. The experimental saturation field is expected to be at $70.7$\,T just above the highest available magnetic field for this experiment. Calculation of the magnetization using the magnetic exchange interaction found in the analysis of the inelastic neutron scattering (as described in the section 'Magnetic excitations') is also shown in the figure as a black dashed line. This value provides a much more accurate description of the magnetization in the whole magnetic field range. 

\subsection{\emph{Ab initio} calculations}

\begin{figure}[htb!]
\includegraphics[width=0.45\textwidth]{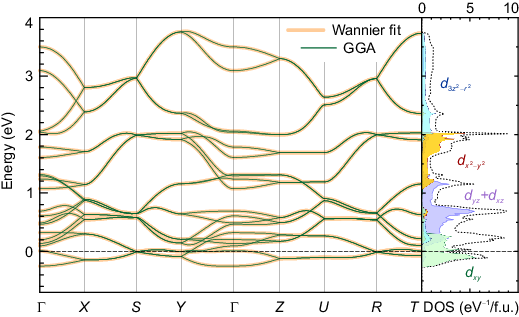}
\caption{GGA bands and their Wannier fit for extracting the hopping parameters. 
The right panel shows orbital-resolved contributions to the density of states and indicates the predominance of the $d_{xy}$ states near the Fermi level. The $k$-path is defined as $\Gamma(0,0,0)$, $X(\frac12,0,0)$, $S(\frac12,\frac12,0)$, $Y(0,\frac12,0)$, $Z(0,0,\frac12)$, $U(\frac12,0,\frac12)$, $R(\frac12,\frac12,\frac12)$, and $T(0,\frac12,\frac12)$.
}
\label{fig:bands}
\end{figure}

To  see  which  of  the  many  possible  exchange terms may be relevant in the case of \vso,  we first perform \emph{ab initio} calculations, where we consider magnetic exchange couplings $J_{ij}$ entering the spin Hamiltonian:

\begin{align} \label{eq:hamiltonian}
\mathcal{H}=  \sum_{\left\langle ij \right\rangle} J_{ij}S_iS_j. \end{align} 

\noindent with the summation running over lattice bonds $\left\langle ij \right\rangle$.
The exchange parameters $J_{ij}$ were extracted using two complementary approaches. On one hand, we calculate hopping integrals between the V $3d$ states using Wannier functions constructed for the uncorrelated (GGA) band structure, and introduce these hoppings into the Kugel-Khomskii model that then delivers the magnetic exchange couplings as follows \cite{mazurenko2006, tsirlin2011},

\begin{align} \label{eq:KKexchange}
J_{ij}= \frac{(4t_{ij}^{(nn)})^2}{U_{ {\rm eff}}}
- \sum_m  \frac{4(t_{ij}^{(nm)})^2 J_{ {\rm eff}}}{(U_{ {\rm eff}} + \Delta_m)(U_{ {\rm eff}}-J_{ {\rm eff}} + \Delta_m)}
\end{align}

\noindent where the first and second terms stand for the antiferromagnetic ($J_{ij}^{AFM}$) and ferromagnetic ($J_{ij}^{FM}$) contributions, respectively. Here, $U_{\mathrm{eff}} = 4$\,eV  is the effective on-site Coulomb repulsion and $J_{\mathrm{eff}} = 1$\,eV is the effective Hund’s coupling in the V $3d$ shell \cite{tsirlin2011, tsirlin2011b}. The hoppings $t_{ij}^{(nn)}$ are between the half-filled states of vanadium, whereas $t_{ij}^{(nm)}$ involve the empty states $m$, and $\Delta _m = E_m-E_n$ is the crystal-field splitting. 
GGA bands and their Wannier fit are shown in Fig. \ref{fig:bands}. 
Alternatively, we obtain the exchange couplings by a mapping procedure \cite{xiang2011, tsirlin2014} using total energies of collinear spin configurations evaluated within DFT+U, where correlation effects in the V $3d$ shell are treated on the mean field level with the on-site Coulomb repulsion $U_d = 4$\,eV,  Hund’s coupling $J_d = 1$\,eV, and atomic-limit flavor of the double-counting correction \cite{saul2014, weickert2016}. 

\begin{figure}[htb!]
\includegraphics[width=0.45\textwidth]{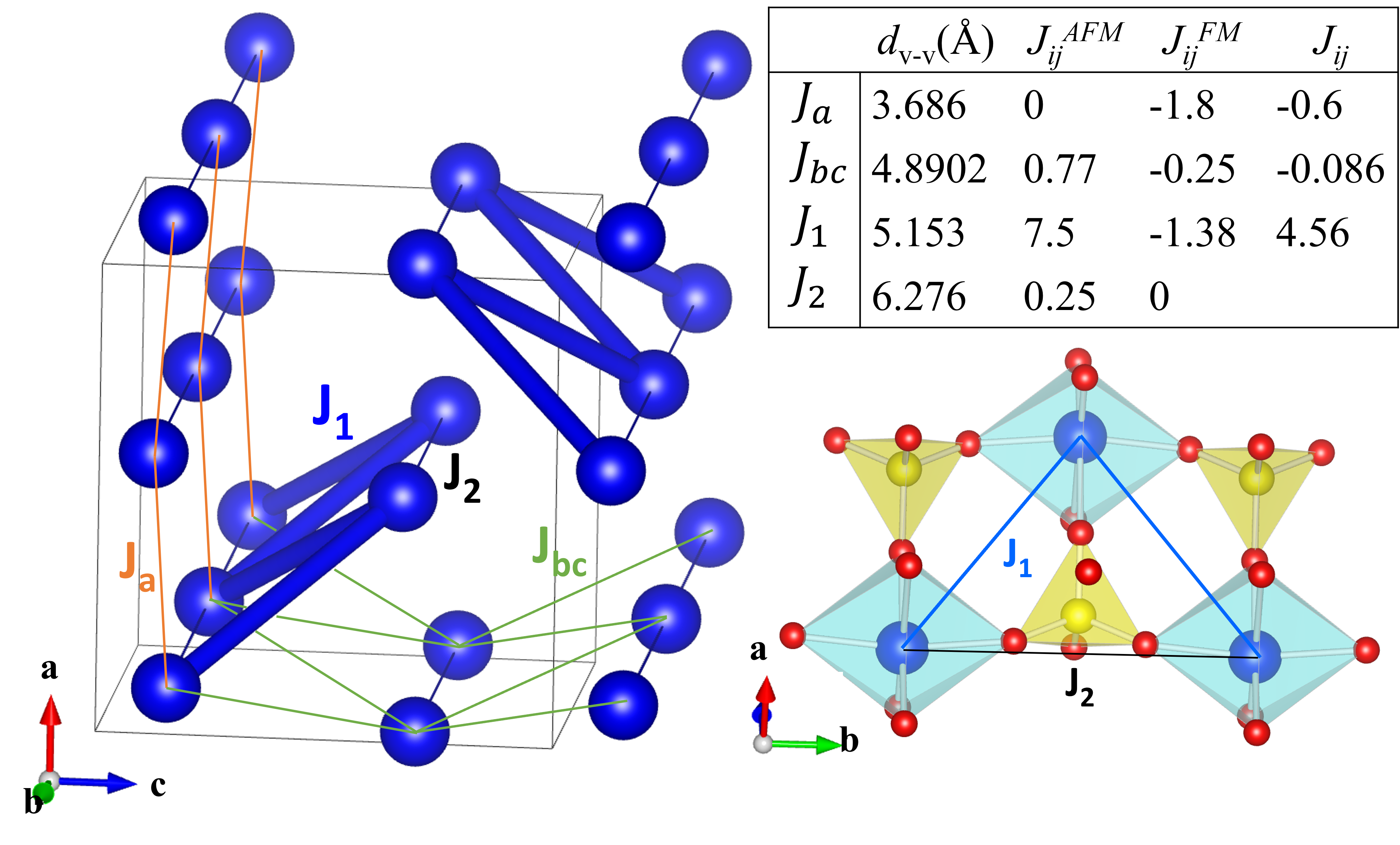}
\caption{Exchange couplings in \vso. The crystal representations were created with \texttt{Vesta} \cite{vesta}. Blue balls represent vanadium ions, yellow is sulfur and red is oxygen. The values of $J_{ij}^{AFM}$ and $J_{ij}^{FM}$ are listed in the inserted table in meV units. These are derived from the Kugel-Khomskii model, eq.~(\ref{eq:KKexchange}), and show relative contributions of different superexchange mechanisms. Total exchange couplings are $J_{ij}$ obtained by the DFT+U mapping analysis and may be different from $J^{AFM}$.}
\label{fig:strucJs}
\end{figure}

In the absence of correlations, the GGA band structure of \vso \,  is metallic and features a broad V $3d$ band that spans the energy range between $-0.3$ and $3.8$\,eV. The lower part of this band is dominated by the half-filled $d_{xy}$ states. Here, we define the local coordinate system in such a way that the $z$-axis points along the short (vanadyl) bond of the VO$_{6}$ octahedron, whereas $x$ and $y$ point along the longer V–O bonds that are approximately perpendicular to it. Then, $z$ is close to the $a$ direction and thus nearly parallel to the structural chains in \vso. Only the $d_{xy}$ states are half-filled, and all other $3d$ states of vanadium are empty, as typical for V$^{4+}$ \cite{mazurenko2006, tsirlin2014, korotin1999}.

Both ferro- and antiferromagnetic contributions to the exchange couplings obtained from eq.\ref{eq:KKexchange}, as well as total exchange couplings $J_{ij}$ obtained by the mapping procedure, are listed and represented graphically in Fig.~\ref{fig:strucJs}. Long-range interaction terms not included in this table are well below 0.008\,meV and can be safely neglected. The nearest-neighbour coupling along the structural chains ($J_a$) is relatively weak and ferromagnetic, while the strongest coupling, $J_1$ $\approx4.3$\,meV, runs between the VO$_{6}$ octahedra that are linked via two SO$_{4}$ tetrahedra (see Fig.~\ref{fig:strucJs}). Such a coupling regime is common for V$^{4+}$ phosphates, arsenates, and other compounds with tetrahedral polyanions \cite{tsirlin2011c, arjun2019}. On the one hand, double tetrahedral bridges create a V–O. . .O–V superexchange pathway that are reasonably efficient despite the large V–V separation of more than 5\,$\rm{\AA}$. On the other hand, the position of the half-filled $d_{xy}$ orbital in the plane perpendicular to the structural chain prevents the $xy$-$xy$ hopping (and, thus, an antiferromagnetic interaction) for the shorter V–V separation along the chain and renders $J_a$ weakly ferromagnetic \cite{tsirlin2011c, nath2008}. The remaining couplings $J_2$ and $J_{bc}$ are much weaker as each of them is mediated by single SO$_{4}$ tetrahedron with a less efficient orbital overlap \cite{roca1998}. Orbital directions can be depicted similarly to those reported in Ref.\cite{tsirlin2011}.

The DFT+U calculations described above were performed for the supercells doubled along the $b$ and $c$ directions without including the distortion reported in Ref. \cite{Fuller}. To account for the effect of structural disorder, we considered two possible configurations, ++++ and +-+- in the notation of Ref. \cite{Fuller}.  Crystal structures of both configurations were optimized on the GGA level, and exchange couplings were calculated by the DFT+U mapping procedure. We found $J_1=3.53$\,meV and $J_a=-0.6$\,meV for the ++++ configuration vs. $J_1=3.7$\,meV and $J_a=-0.77$\,meV for the +-+- configuration. These very minute changes suggest that structural disorder has a minor effect on magnetism.

\subsection{Magnetic excitations}

\begin{figure*}[t]
\includegraphics[width=1.05\textwidth]{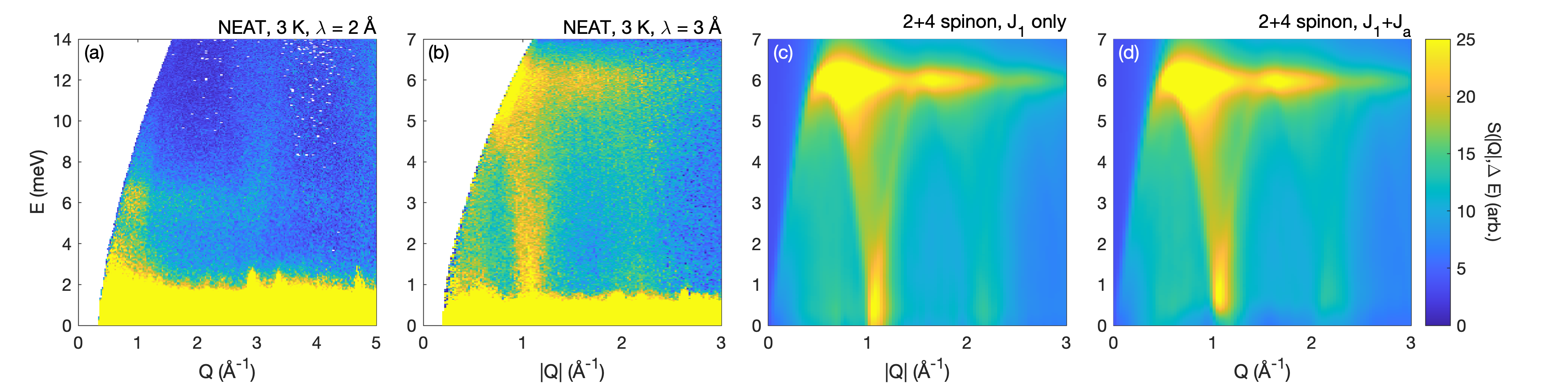} \\
\includegraphics[width=1.05\textwidth]{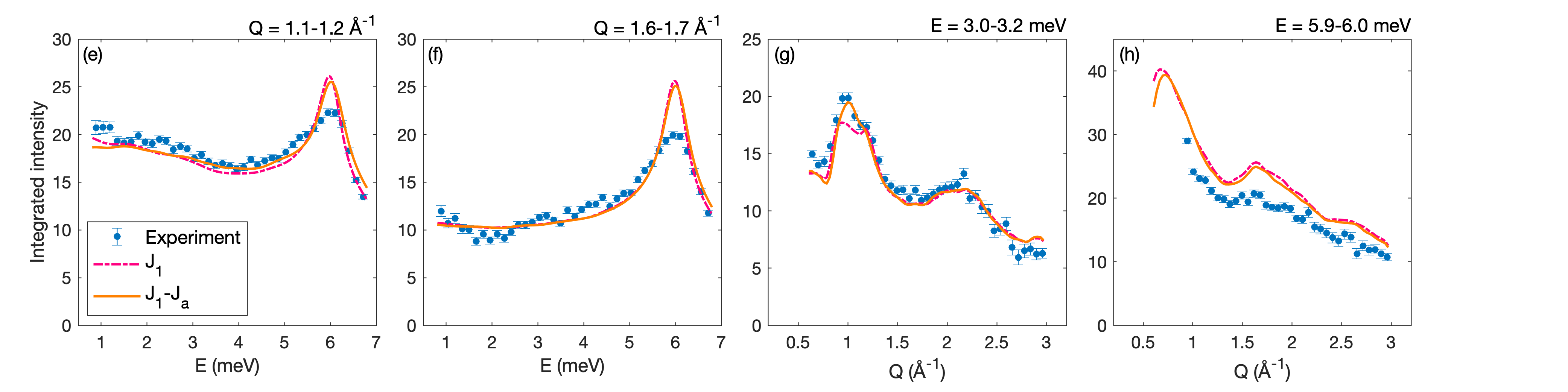}
\caption{(a,b) Experimental $S(Q,\Delta E)$ for \vso \, at $3$K measured at neutron wavelengths $\lambda = 2$\,$\rm{\AA}${} and $3$\,$\rm{\AA}${} on the NEAT spectrometer. (c,d) Calculated $S(Q,\Delta E)$ for the $J_1$-only (isolated chain, c) and $J_1+J_a$ (d) models. The model parameters are given in the text. (e-h) Constant-Q (e,f) and $\Delta E$ (g,h) cuts through the experimental $3$\,$\rm{\AA}$ \, 
data. Corresponding cuts through the calculated $S(Q,\Delta E)$ are indicated by red dot-dashed ($J_1$ only) and orange solid ($J_1-J_a$) lines. The intensity is overestimated by around $20\%$ at $6$~meV (h).}
\label{fig:inelastic_neutron}
\end{figure*}

The dynamical structure factors $S(|Q|,\Delta E)$ of \vso \, collected at $3$~K with neutron wavelengths of $2$ and $3$~$\rm{\AA}$~($E_i = 20.45,~9.09$~meV respectively) are shown in Figures \ref{fig:inelastic_neutron}(a) and (b). Magnetic scattering is clearly evident at low $|Q|$, dispersing from $|Q_0|\sim 1.05$~$\rm{\AA}$$^{-1}$ and extending in energy up to around $10$~meV [Fig.~\ref{fig:inelastic_neutron}(a)], with a sharp peak at $\Delta E \sim 6$~meV. Weaker magnetic intensity is also observed near the elastic line around $|Q| \sim 0.6$ and $2$~$\rm{\AA}$$^{-1}$. Overall, the spectrum is reminiscent of powder data from other one-dimensional systems, like KTi(SO$_4$)$_2$ \cite{Nilsen2015},  Sr$_3$CuPtO$_6$ \cite{Leiner2018} and KMoOP$_2$O$_7$ \cite{ KMoOP2O7}. The observed $|Q_0|$ is close to the double-(SO$_4$)$^{2-}$-bridged zig-zag chains that run along the $b$ axis [Fig.~\ref{fig:strucJs}] in agreement with the $ab$-initio calculations. 

To further confirm the hypothesis that the experimental data can be reproduced assuming one-dimensional zig-zag chains along $b$, we compare our experimental data to the total dynamical structure factor $S_{1d}(Q_y,\Delta E)$ for the 2- and 4-spinon continua calculated using the Bethe ansatz \cite{Caux2006}. To do this, the theoretical $S_{1d}(Q,\Delta E)$ is first recalculated due to the strong deviation of the chains from linearity and the resulting folding of the Brillouin zone. From the orthorhombic symmetry of the material, the folded $S(\mathbf{Q},\Delta E)$ is expressed as:

\begin{align}
S(\mathbf{Q},&\Delta E) =
\cos^2{(Q_x\delta_x)}\cos^2{(Q_z\delta_z)}S_{1d}(\mathbf{Q},\Delta E) \nonumber \\ 
+& \cos^2{(Q_x\delta_x)}\sin^2{(Q_z\delta_z)}S_{1d}(\mathbf{Q}+(110),\Delta E) \nonumber  \\
+& \sin^2{(Q_x\delta_x)}\cos^2{(Q_z\delta_z)}S_{1d}{(\mathbf{Q}+(111)},\Delta E) \nonumber \\
+& \sin^2{(Q_x\delta_x)}\sin^2{(Q_z\delta_z)}S_{1d}(\mathbf{Q}+(011),\Delta E).
\label{bz}
\end{align}

\noindent Here, $S_{1d}(\mathbf{Q},\Delta E)$ represents the linear chain 2- and 4-spinon dynamical structure factor shifted by a $\mathbf{Q}$-dependent lattice vector from the antiferromagnetic zone center $(010)$, and $\delta_x = 0.1615$ and $\delta_z = 0.2353$ are the offsets of the V$^{4+}$ position $(\delta_x,\frac{1}{4},\delta_z)$ from that which would generate a perfectly linear chain: $(0,\frac{1}{4},0)$. Following folding, the theoretical data is powder averaged, yielding $S(|Q|,\Delta E)$, shown in Fig.~\ref{fig:inelastic_neutron}(c). The correct zone-boundary energy, corresponding to the maximum in $\Delta E$ at $6$~meV, is recovered for an intra-chain coupling $J_1 = 3.83(2)$~meV, and most details of the spectrum, including the weak intensities away from the main branch of the continuum are also reproduced ($\chi^2 =\sum_{(Yobs-Ycalc)^2}= 5.3$) for 6 constant-$|Q|$ cuts through $S(|Q|,\Delta E)$. On the other hand, the intensity at the zone-boundary energy is somewhat overestimated. This reduction in intensity is consistent with the presence of inter-chain couplings, as seen, for example, in the case of Cs$_2$Cu$X_4$ ($X$=Cl, Br) \cite{Coldea2003}. 
This single $J_1$ value was used to calculate the magnetization using the Fisher and Bonner expression as previously described. As neutron scattering measurements are not affected by the g-factor, the ESR literature g factor of 1.987 was used for this calculation \cite{EPR}. Fig. \ref{fig:magn} shows the calculation as a black dashed line, again no fitting procedure was used. This comparison shows that a single magnetic intra-chain coupling of $J_1 = 3.83(2)$~meV  is a highly accurate value for \vso.

The predictions from our \textit{\emph{Ab initio}} calculations suggest that the dominant exchange beyond $J_1$ is $J_a$, while $J_{bc}$, and $J_2$ are somewhat smaller. Therefore, we performed fits of the dynamical structure factor $S(|Q|,\Delta E)$ for the $J_1$-$J_a$ model. This was done following the random phase approximation-style approach in \cite{Kohno2007}, from which $S(\mathbf{Q},\Delta E)$ is given by the following expression:

\begin{align}
\frac{S_{1d}(Q_y,\Delta E)}{\left[1+J^{\prime}(\mathbf{Q})\chi^{\prime}(Q_y,\Delta E)\right]^2+\left[J^{\prime}(\mathbf{Q})\chi^{\prime\prime}(Q_y,\Delta E)\right]^2},
\label{eq:rpa}
\end{align}

\noindent where $S_{1d}(Q_y,\Delta E)$ is defined as above and

\begin{align}
J^{\prime}(\mathbf{Q})&\simeq J_a \cos{(Q_x/2)}\cos{(Q_z/2)}, 
\end{align}

\noindent where the Fourier components for the further neighbour couplings correspond to that of the straight chain.  It is necessary to normalize $S_{1d}(Q_y,\Delta E)$ such that $\int_{0}^{2\pi}{\int_{0}^{\infty} {S_{1d}(k_y,\Delta E)}}d\Delta E dk_{y} = 1$ for the exchanges extracted from (\ref{eq:rpa}) to be physically meaningful. The imaginary and real dynamical susceptibilities $\chi^{\prime\prime}(Q_y,\Delta E)$ and $\chi^{\prime}(Q_y,\Delta E)$ were calculated using $S_{1d}(Q_y,\Delta E)=\chi^{\prime\prime}(k_y, \Delta E)/\pi$ and a numerical Kramers-Kronig transformation, respectively. Finally, Brillouin zone folding was accounted for before powder averaging to obtain $S(|Q|,\Delta E)$. Due to the time-consuming nature of this procedure, the fitting was performed using six constant-$|Q|$ cuts through the data rather than the full data set. The optimization was carried out using the particle swarm method, and a single constant was added to all cuts to account for the instrumental background. The fit to the $J_1-J_a$ model yielded a slight reduction in $\chi^2$ from $5.3$ (for the $J_1$ only model) to $5.1$, and a small $J_a = 0.2(1)$meV and nearly unchanged $J_1=3.81(2)$meV [Fig.~{\ref{fig:inelastic_neutron}(d-h)}]. The inclusion of terms beyond $J_a$ did not result in any further improvements to the fit. Dzyaloshinskii–Moriya interactions could be further considered in the model as these are allowed due to the $2\%$ crystallographic disorder and this is the scope of future work.

\section{Conclusions}

Our results show that the spin chains in \vso \, are formed by
doubly (SO$_{4}$)$^{2-}$-bridged chains along the $b$-direction that do not coincide with the structural chains of the V$^{4+}$O$_{6}$ octahedra. These spin chains are magnetically coupled by the magnetic exchange interaction $J_1=3.83$\,meV.  This is a common occurrence in V$^{4+}$ compounds due to the nature of their half-filled magnetic orbital \cite{GarrettAW1997Meit, TennantDA1997ESaS, WeickertFranziska2019Fdda}. 
$J_2$ is the second neighbour coupling along the spin chain and 
it is at least one order of magnitude weaker than $J_1$. It may be responsible for a weak frustration therein, although it is by far insufficient to open a spin gap, which was indeed not detected in any of our measurements. The coupling $J_a$ links the spin chains into a three-dimensional network but with low connectivity as there are two such couplings per vanadium site, as opposed to the four inter-chain couplings per site assumed in a standard quasi 1D -antiferromagnet. The remaining discrepancy between the modelled INS intensity and the experimental data at $\sim$6~meV likely has a different origin, the investigation of which must await the availability of larger single crystals.
The weak connectivity of the spin chains suggests that the N\'eel temperature of \vso \, could be quite low. Indeed, we do not detect any magnetic order down to $0.085$\,K within the used experimental settings.
We do not detect any effect of the reported crystallographic disorder in the magnetic properties of \vso, other than possibly a finite amount of paramagnetic impurities.

\section{Acknowledgments}
We thank M. Reehius, B. Lake, D. Chernyshov, and C. Fuller for fruitful discussions as well as P. Karen for ini- 
tial XRD results. Additional thanks to Helmholtz Zentrum Berlin for supporting A.B.-C. during this project through the 
summer student program. This work is based on experiments performed at the research reactor BER-II at the Helmholtz 
Zentrum Berlin, Germany, ISIS neutron and muon source in Rutherford Appleton Laboratory and Swiss-Norwegian 
Beamlines, at the ESRF. Collaboration with the Solid State Insitute at the University of Tokyo was possible thanks 
to travel grant YYF given by the University of Stavanger to D. Q.-C..

\end{document}